\documentclass[%
 aip,
 amsmath,amssymb,
 reprint,%
]{revtex4-1}

\usepackage{graphicx}
\usepackage{dcolumn}
\usepackage{bm}
\usepackage[utf8]{inputenc}
\usepackage[T1]{fontenc}
\usepackage{mathptmx}
\usepackage{etoolbox}
\usepackage{amsmath,amssymb,amsfonts}
\usepackage{amsthm}
\usepackage{acronym}
\usepackage{xcolor}
\usepackage{physics}
\usepackage{hyperref}
\usepackage[capitalise]{cleveref}
\usepackage{caption}
\usepackage{subcaption}
\usepackage{ulem}
\usepackage{booktabs}
\usepackage{url}

\definecolor{modify}{RGB}{0,0,255}%

\makeatletter \def\switch@array{}\makeatother

\makeatletter
\def\@email#1#2{%
 \endgroup
 \patchcmd{\titleblock@produce}
  {\frontmatter@RRAPformat}
  {\frontmatter@RRAPformat{\produce@RRAP{*#1\href{mailto:#2}{#2}}}\frontmatter@RRAPformat}
  {}{}
}%
\makeatother

\newcommand{\ve}[1]{\boldsymbol{#1}}
\begin{document}


\newacro{qimr}[QImR]{Quantum Image Representation}
\newacro{qip}[QIP]{Quantum Image Processing}
\newacro{tnr}[TNR]{Tensor Network Representation}
\newacro{frqi}[FRQI]{Flexible Representation of Quantum Images}
\newacro{efrqi}[EFRQI]{Enhanced Flexible Representation of Quantum Images}
\newacro{neqr}[NEQR]{Novel Enhanced Quantum Representation}
\newacro{eneqr}[ENEQR]{Enhanced Novel Enhanced Quantum Representation}
\newacro{qpie}[QPIE]{Quantum Probability Image Encoding}
\newacro{tn}[TN]{Tensor Network}
\newacro{mps}[MPS]{Matrix Product State}

\newacro{ml}[ML]{Machine Learning}
\newacro{nn}[NN]{Neural Network}
\newacro{svm}[SVM]{Support Vector Machine}

\newacro{qml}[QML]{Quantum Machine Learning}
\newacro{nisq}[NISQ]{Noisy Intermediate-Scale Quantum}
\newacro{qpu}[QPU]{Quantum Processing Unit}

\preprint{AIP/123-QED}

\title{Analysis of Quantum Image Representations for Supervised Classification}
\author{Marco Parigi}%
\affiliation{Department of Physics and Astronomy, University of Florence, Via Sansone 1, Sesto Fiorentino, 50019, Florence, Italy.}%
 
\author{Mehran Khosrojerdi}%
\email[Corresponding Author: ]{mehran.khosrojerdi@unifi.it}
\affiliation{Department of Physics and Astronomy, University of Florence, Via Sansone 1, Sesto Fiorentino, 50019, Florence, Italy.}%

\author{Filippo Caruso}
\affiliation{Department of Physics and Astronomy, University of Florence, Via Sansone 1, Sesto Fiorentino, 50019, Florence, Italy.}%

\author{Leonardo Banchi}
\affiliation{Department of Physics and Astronomy, University of Florence, Via Sansone 1, Sesto Fiorentino, 50019, Florence, Italy.}%
\affiliation{INFN Sezione di Firenze, via G. Sansone 1, I-50019, Sesto Fiorentino (FI), Italy}

\date{\today}

\begin{abstract}
In the era of big data and artificial intelligence, the increasing volume of data and the demand to solve more and more complex computational challenges are two driving forces for improving the efficiency of data storage, processing and analysis. Quantum image processing (QIP) is an interdisciplinary field between quantum information science and image processing, which has the potential to alleviate some of these challenges by leveraging the power of quantum computing. In this work, we compare and examine the compression properties of four different Quantum Image Representations (QImRs): namely, Tensor Network Representation (TNR), Flexible Representation of Quantum Image (FRQI), Novel Enhanced Quantum Representation NEQR, and Quantum Probability Image Encoding (QPIE). Our simulations show that FRQI and QPIE  perform a higher compression of image information than TNR and NEQR. Furthermore, we investigate the trade-off between accuracy and memory in binary classification problems, evaluating the performance of quantum kernels based on QImRs compared to the classical linear kernel. Our results indicate that quantum kernels provide comparable classification average accuracy but require exponentially fewer resources for image storage.
\end{abstract}

\maketitle

\section{Introduction}

\ac{qip} is a research field of quantum information science that attempts to overcome the limitations of classical computers, harnessing the peculiar properties of quantum mechanical systems, e.g. quantum superposition and entanglement, 
to provide a more efficient way to store, manipulate, and extract visual information from digital images~\cite{Yan2016, Yao2017_qpie}. 

On a classical computer, a black and white digital image with $N = 2^{n} \times 2^{n}$ pixels is represented as a matrix and encoded by $N$ bits. In contrast, the number of qubits needed to store the same image on a quantum device can be $\mathcal O(\log N)$. We therefore obtain an exponential reduction in computational space resources to store images on quantum processors~\cite{Yao2017_qpie}. Furthermore, quantum operations~\cite{ Nielsen_Chuang_2010, Weinstein_1DQFT_2001} such as Fourier, Hadamard, and Haar wavelet transforms, which are usually included as subroutines in image processing tasks, provide exponential speed-up over their classical counterparts~\cite{Hoyer1997efficientquantumtransforms, Fijany_Q_Wavelet_Transforms1999, Yao2017_qpie}, which is nonetheless challenged by state preparation costs~\cite{aaronson2015read}.

Reduced storage requirements can also provide higher accuracies in machine learning applications. Indeed, there is a tight connection between learning and compression, which has been explored in both classical deep learning \cite{tishby2015deep} and quantum learning settings \cite{banchi2021generalization,banchi2024statistical}. 
This connection is quite intuitive: even humans learn new topics not just by merely memorizing all the ``training data''  but rather by extracting and compressing \textit{relevant} information from training sources, such as books. 
By exploiting the superposition principle, quantum states can compress classical data, not only images \cite{elliott2020extreme}, with exponentially  fewer resources without  loss of accuracy. Moreover, the advantage still persists when some loss or noise is tolerated, paving the way to machine learning applications \cite{banchi2024accuracy, yang2025dimension}. Recent studies have demonstrated that hybrid quantum–classical architectures can leverage these compression principles for image classification\cite{Slabbert2025}. 

The \ac{qimr} is a subarea of \ac{qip} that focuses on image processing tasks and how well they can be performed on quantum hardware.
In a classical computer, digital images are defined as matrices of
numbers representing the discrete color or intensity values
present in every image pixel. In other words, an image is an object that carries two pieces of information: the position and the color of each pixel.
Over the years, many \acp{qimr} have been proposed that employ different models and techniques to encode the intensity and position information of pixels~\cite{Yan2016, Lisnichenko2022}.

A general quantum state that encodes the color and the position of each pixel in a digital image $I$ has the following form:
\begin{equation}\label{eq:general_form_qimagestate}
    \ket{I} = \frac{1}{\sqrt{N}} \sum_{P=0}^{N-1} \ket{C_P} \otimes \ket{P},
\end{equation}
where $N$ is the number of pixels and the two quantum states, $\ket{C_P}$ and $\ket{P}$, respectively, encode the color and position of a pixel.
Various models for representing quantum images have been investigated and implemented on \ac{nisq} devices. Das et al.~\cite{Das2023} notably realized a quantum pattern recognition protocol on a  real \ac{qpu}\cite{qpu} choosing the \ac{qpie}~\cite{Yao2017_qpie}. Furthermore, Geng et al.~\cite{Geng2023} explored the possibilities of implementing \acp{qimr} on superconducting and trapped-ion quantum computers, and successfully implemented a $2\times 2$ \ac{frqi}~\cite{Le2011_frqi} image. 

In addition to these approaches, tensor-network-based quantum image representations have shown strong potential for compressing and representing image data in a way that is naturally compatible with quantum circuit architectures. In fact, Stoudenmire et al.~\cite{Stoudenmire2016} demonstrated how quantum-inspired tensor networks can be applied effectively to supervised learning tasks, especially image classification, by representing data with reduced dimensionality while preserving key features.

In this work, we consider four representations for grayscale images that incorporate classical image information differently, namely: i) \ac{tnr}~\cite{latorre2005_realket}, ii) \ac{frqi}~\cite{Le2011_frqi}, iii) \ac{neqr}~\cite{Zhang2013_neqr} and iv) \ac{qpie}~\cite{Yao2017_qpie}. Table~\ref{tab: comparing-qimrs} summarizes the most important parameters and characteristics of the four encoding methods.

The rest of this article is organized as follows. We first describe how these representations encode the color and position of pixels, discussing their advantages and drawbacks. Then we investigate the trade-off between accuracy and memory in binary classification and finally discuss a few applications for learning to classify images.

\begin{table*}[ht]
\centering
\renewcommand{\arraystretch}{1.1}
\begin{tabular*}{\textwidth}{@{\extracolsep{\fill}}p{3.5cm}p{2.5cm}p{2.5cm}p{3cm}p{3cm}}
\toprule
\textbf{Quantum Image} & \textbf{Color} & \textbf{Qubit} & \textbf{Complexity} & \textbf{Retrieval} \\
\textbf{Representation} & \textbf{encoding} & \textbf{resource} & & \\
\midrule
\ac{tnr} & Tensor & $2n + 1$ & -- & Probabilistic \\
\ac{frqi} & $1$ Angle & $2n + 1$ & $\mathcal{O}(2^{4n})$ & Probabilistic \\ 
\ac{neqr} & Basis States & $2n + q$ & $\mathcal{O}(qn2^{2n})$ & Deterministic\\ 
\ac{qpie} & Amplitudes & $2n$ & $\mathcal{O}(2^{n})$ & Probabilistic\\ 
\bottomrule
\end{tabular*}
\caption{Comparison of key parameters and characteristics of four quantum image representations, \ac{tnr}, \ac{frqi}, \ac{neqr}, and \ac{qpie}, for encoding images of size $2^n \times 2^n$ with grayscale values $2^q$.}
\label{tab: comparing-qimrs}
\end{table*}

\section{Quantum Image Processing}\label{sec:qip}
\subsection{Tensor Network Representation}
\begin{figure}[t]
	\begin{center}
	\includegraphics[width=0.45\textwidth, trim=0 20 0 0, clip]{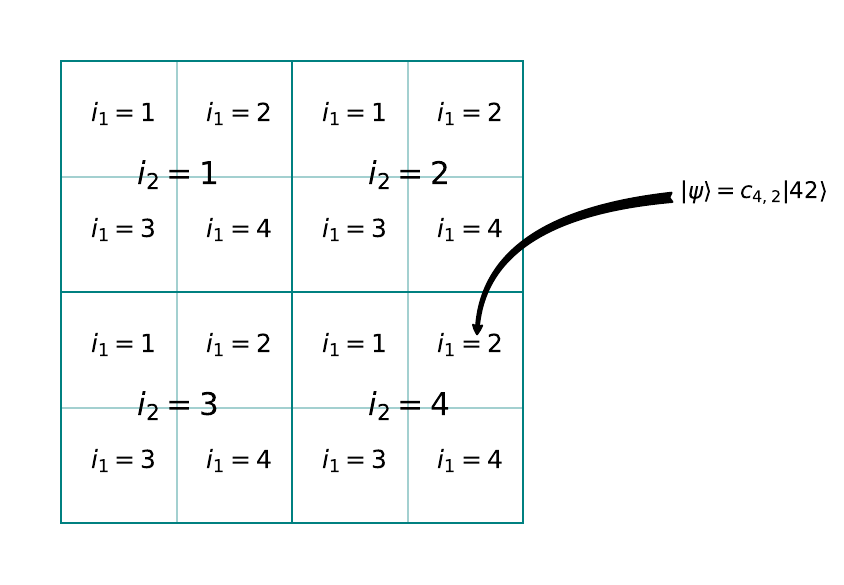}
	\end{center}
  \caption{Example real-ket indexing for an image with $2^2 \times 2^2$ pixels}
	\label{fig:realket_example}
\end{figure}

In quantum mechanics, a \textit{ket} is a mathematical construct used to represent states in a Hilbert space. This concept can be applied to images by mapping pixel values into a structured quantum format. Using a block-structured addressing method, pixels of an image can be organized in a way that reflects their spatial relationships, allowing them to be represented as a quantum state with real amplitudes, here called \textit{real-kets}. This transformation provides a foundation for analyzing images using quantum-inspired techniques.
Each pixel in an image can be represented using a certain number of classical bits. For instance, in an  $8$-bit grayscale image, each pixel is  an integer between 0 (black) and  255 (white). 
An image with $2^n\times 2^n$ pixels can be mathematically represented as a real-ket as follows~\cite{latorre2005_realket}

\begin{equation}
    \ket{\psi_{2^n \times 2^n}} = \sum_{i_1=1}^4\cdots\sum_{i_n=1}^4  c_{i_n,\cdots,i_1} \ket{i_n \cdots i_1},
    \label{eq:realket}
\end{equation}
where $c_{i_n,\cdots,i_1}$ stores the pixel values, demonstrating how the image is encoded in a structured quantum state. As an example for better understanding is shown in  \cref{fig:realket_example} for $n=2$. In other terms, an image is iteratively reduced in four larger blocks, each indexed with an integer $i=1,\dots,4$. This reduction is performed $n$ times until each block is composed only of a single pixel. The position of each pixel can hence be expressed using $n$ integers $1\dots4$.  
In the example of  \cref{fig:realket_example}, each sub-block is assigned an index based on its position: $i_2 = 1$ corresponds to the upper-left pixel, $i_2 = 2$ to the upper-right, $i_2 = 3$ to the lower-left, and $i_2 = 4$ to the lower-right. This labeling scheme provides a structured way to organize pixel positions within the block.  
To specify which sub-block is being addressed within the larger structure, we introduce an additional label, $i_1$, following the same convention used for the inner sub-blocks. This hierarchical labeling ensures a consistent and scalable representation of pixel organization. 

\begin{figure}[t]
	\begin{center}
	\includegraphics[width=0.35\textwidth, trim=77 65 55 65, clip]{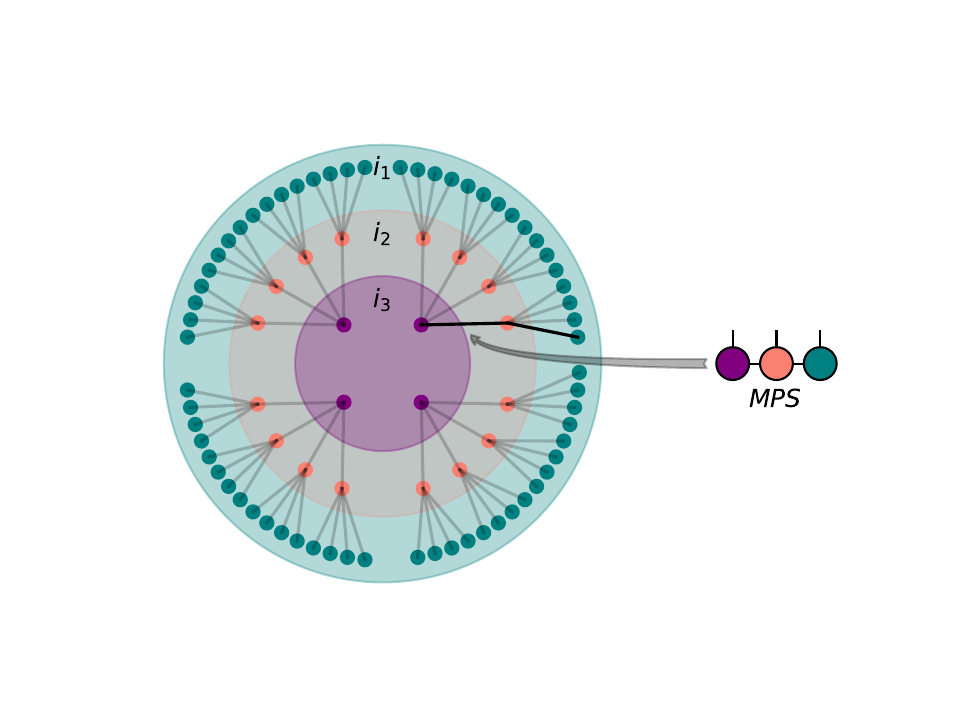}
	\end{center}
  \caption{A typical image in the framework of tensor networks}
	\label{fig:mps_form}
\end{figure}

Images with varying textures and structures exhibit different levels of entanglement, capturing the complexity of pixel correlations. This formulation enables the use of mathematical expressions to describe an image as a quantum state, where the tensors encode the relationships between pixels. Once the image is mapped to a quantum state, a particularly useful framework for compressing it is via \acp{tn} methods or, specifically, \acp{mps}. A \ac{mps} provides a compact way to encode an image by organizing pixel information into an array of tensors that efficiently capture entanglement between different regions. The \ac{mps} representation can be mathematically expressed as
\begin{equation}
  \ket{\psi_{\Gamma}} = 
  \!\!\! 
  \sum_{\{i_j\},\{\lambda_j\}}
  \!\!\! 
    \Gamma_{\lambda_1\lambda_2}^{(1)i_1} \Gamma_{\lambda_2\lambda_3}^{(2)i_2} \cdots \Gamma_{\lambda_n\lambda_1}^{(n)i_n}
    \ket{i_n, \cdots i_2, i_1},
    \label{eq:MPS_realket}
\end{equation}
where $\Gamma^{(j)}$ are tensors with real elements. The advantage of \ac{mps} lies in its ability to represent images using fewer parameters, making it an efficient tool for compression and structured encoding. Moreover, the \ac{mps}  offers a natural bridge between quantum information theory and traditional image processing. Indeed, in the \ac{mps} encoding of an image, the amount of entanglement corresponds to classical pixel correlations. To further illustrate the concept of \ac{mps} at this stage, we refer to  \cref{fig:mps_form}. This figure visually demonstrates how blocks and sub-blocks are interconnected and how a Tree Tensor Network (TTN) in one dimension can be formulated as an \ac{mps} representation, providing a structured approach to encoding pixel relationships of a typical image.

By leveraging real-kets and \acp{mps}, images can be encoded efficiently, and the structured nature of these representations allows for a deeper understanding of how images store information.
Moreover, from the tensors $\Gamma$ in Eq.~\eqref{eq:MPS_realket}, it is possible\cite{rudolph2023synergistic}
to obtain a quantum circuit for creating the same state in a \ac{nisq} device.

\subsection{Flexible Representation of Quantum Images}
\begin{figure}
    \centering
    \includegraphics[width=0.5\linewidth]{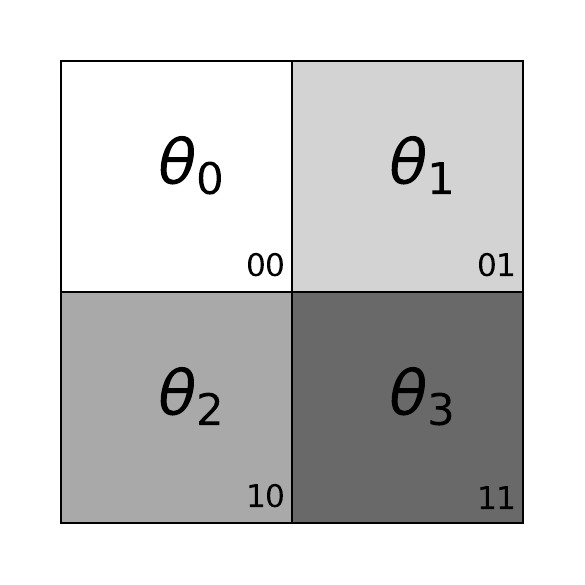}
    \caption{A $2\times 2$ \ac{frqi} image. The angles $\theta_i$, with $i = 0, 1, 2, 3$, at the center of each pixel encode the grays intensity of the corresponding pixels in the angle representation, while the 2-bit strings in the lower-right corner identify the positions in the image in binary representation.}
    \label{fig:frqi_image}
\end{figure}

In 2010, Le et al.~\cite{Le2011_frqi} proposed Flexible Representation of Quantum Images (\ac{frqi}) that uses quantum superposition to encode a classical image in the form of a normalized quantum state. More precisely, inspired by the pixel representation for images in classical computers, the \ac{frqi} encodes a $2^n \times 2^n$ grayscale image in a quantum state as follows:
\begin{equation}\label{eq:frqi_state}
    \ket{I(\theta)} = \frac{1}{2^n} \sum_{i=0}^{2^{2n}-1} \ket{c(\theta_i)} \otimes \ket{i}, 
\end{equation}
where the color-state,
\begin{equation}
    \ket{c(\theta_i)} = \cos\theta_{i} \ket{0} + \sin\theta_{i}\ket{1},
\end{equation}
with $\ket{0}$ and $\ket{1}$ are the two-dimensional quantum states of the computational basis, $\ket{i}$ is a sequence of $2n$ qubits storing the position of the pixel, and $\theta_i \in [0, \frac{\pi}{2}]$ is the angle that encodes the color of the $i$-th pixel. An example of a $2\times 2$ \ac{frqi} image is illustrated in \cref{fig:frqi_image}.
The \ac{frqi} requires $2n + 1$ qubits to represent an image on a quantum device, exponentially reducing the space resources needed in the classical case.  Specifically, $2n$ qubits encode the positions of the pixels, while a single qubit stores the grayscale information of all pixels. These are commonly referred to as position qubits and color qubit, respectively. 
Furthermore, this model provides efficient color representations~\cite{frqi_efficientCT} and fast geometric transformations~\cite{frqi_fastGT}. 
The implementation\cite{Le2011_frqi} of a \ac{frqi} circuit begins by applying Hadamard gates to the position qubits to create a uniform superposition. Then, for each pixel, a multi-controlled rotation gate is applied to the color qubit, where the rotation angle $\theta$ depends on the intensity of the pixel.

The two main \ac{frqi} drawbacks are image retrieval and the depth of the quantum circuit. The first is due to the color encoding of each pixel that is in the probability amplitudes of a qubit, and consequently an infinite number of measurements are necessary to exactly reconstruct the initial image. Therefore, the original classical image can only be retrieved probabilistically. The second significant drawback is that the preparation of a \ac{frqi} state is based on the polynomial preparation theorem~\cite{Le2011_frqi} and has a computational complexity of $\mathcal{O}(2^{4n})$.  However, the last drawback has recently been overcome by Nasr et al.~\cite{Nasr2021} who introduced \ac{efrqi} reducing the computational complexity to $\mathcal{O}(2n2^{2n})$.

\subsection{Novel Enhanced Quantum Representation}
\begin{figure}
    \centering
    \includegraphics[width=0.5\linewidth]{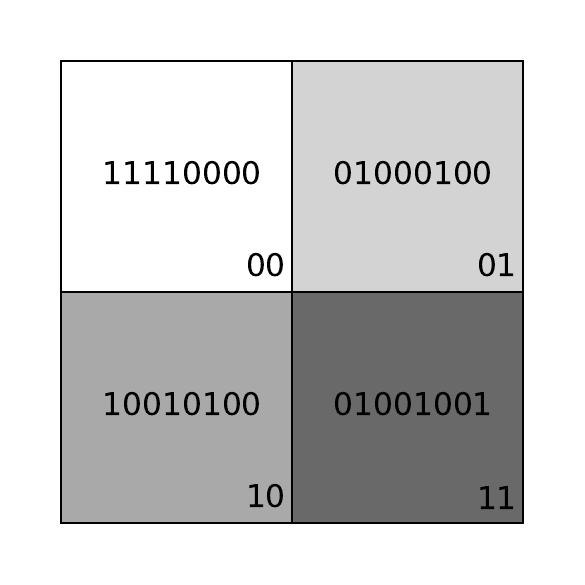}
    \caption{A $ 2\times 2$ \ac{neqr} image. The $8$-bit strings at the center of each pixel and the $2$-bit strings in the lower-right corner encode, respectively, the gray intensity and the positions of the pixels in the image.}
    \label{fig:neqr_image}
\end{figure}

The \ac{neqr} representation has been proposed by Zhang et al.~\cite{Zhang2013_neqr} to overcome the limitations of \ac{frqi}.
More precisely, to improve image retrieval, the \ac{neqr} model encodes the digital image in a superposition of two entangled qubit sequences storing grayscale and position information.
Formally, a $2^n \times 2^n$ image with the intensity range $2^q$ (where $q$ equals $8$ for 256 intensity gray levels) is stored in a \ac{neqr} state of the form:
\begin{equation}\label{eq:neqr_state}
\ket{I} = \frac{1}{2^n} \sum_{y=0}^{2^{n}-1}\sum_{x=0}^{2^{n}-1} \ket{f(y,x)} \otimes \ket{yx},
\end{equation}
where $f(y,x) \in [0, 2^{q}-1]$ is a $q$-bit sequence $C^0_{yx} \ldots C^{q-2}_{yx}C^{q-1}_{yx}$ encoding the grayscale value of the $yx$ pixel:
\begin{equation*}
	f(y,x)=C^0_{yx},C^1_{yx},...C^{q-2}_{yx}C^{q-1}_{yx},\quad 
	C^k_{yx}\in[0,1],
\end{equation*} 
 and position states, 
 \begin{equation*}
 	\ket{yx}=\ket{y_0y_1...y_{n-1}} \otimes \ket{x_0x_1...x_{n-1}}.
 \end{equation*}
The \ac{neqr} requires $2n + q$ qubits: $2n$ qubits are used to encode the pixel positions and $q$ qubits to store the grayscale intensity values. An example of a $2\times 2$ \ac{neqr} image is shown in \cref{fig:neqr_image}.
Because \ac{neqr} uses different basis states of qubits, the original classical image can be accurately retrieved through quantum measurements. In addition, some complex color operations, such as partial color operations, can be performed~\cite{Zhang2013_neqr}.
Moreover, \ac{neqr} achieves a quadratic speedup in quantum image preparation with respect to \ac{frqi}, taking a computational complexity of $\mathcal{O}(2qn2^{2n})$. The implementation\cite{Zhang2013_neqr} of a \ac{neqr} circuit begins by applying Hadamard gates to the position qubits, creating a uniform superposition of all pixel positions. Then, multi-controlled NOT quantum gates are applied to encode the grayscale intensity information into the $q$ color qubits, conditioned on the corresponding pixel positions. Recently, \ac{eneqr} has been proposed by Nasr et al.~\cite{Nasr2021} that requires computational complexity $\mathcal{O}(2n2^{2n})$.

\subsection{Quantum Probability Image Encoding}

\begin{figure}
  \begin{center}
    \includegraphics[width=0.5\linewidth]{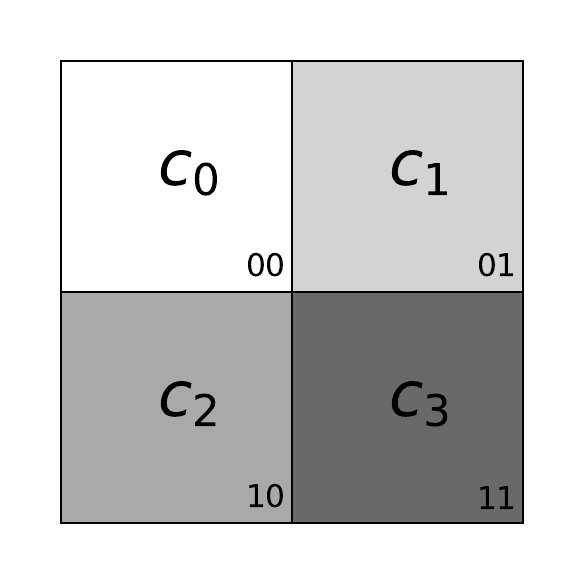}
  \end{center}
  \caption{A $ 2\times 2$ \ac{qpie} image. The amplitudes $c_i$, with $i = 0, 1, 2, 3$, at the center of each pixel encode the grays intensity of the corresponding pixels, while the 2-bit strings in the lower-right corner label the pixel positions in the image in binary representation.}
  \label{fig:qpie_image}
\end{figure}

Quantum Probability Image Encoding (\ac{qpie}), proposed by Yao et al.~\cite{Yao2017_qpie} uses the probability amplitudes of pure quantum states to store the grayscale values of pixels. Formally, given a $2^n \times 2^n$ grayscale digital image, the \ac{qpie} state is the following:
\begin{equation}\label{eq:qpie_state}
\ket{I} = \sum_{i=0}^{2^{2n}-1} c_{i} \ket{i},  \qquad c_i = \frac{I_i}{\sqrt{\sum{I_i}^2}},
\end{equation}
where $I_i$ is the color intensity of  $i$-th pixel, and $c_i$ is the normalized intensity so that the squared sum of all the
probabilities amplitudes is equal to $1$. An example of the \ac{qpie} image is illustrated in \cref{fig:qpie_image}.
For the representation on a quantum computer  the \ac{qpie} only uses $2n$ qubits. This model therefore requires the least number of qubits compared to the \ac{frqi} and \ac{neqr}.
However, because pixel values are stored in the state amplitude, 
\ac{qpie} has the same drawback as \ac{frqi} about retrieval of the original image from a finite number of measurements.
The preparation of a $n$-qubit pure quantum state to a specific probability distribution computational complexity $\mathcal{O}(2^n)$~\cite{Ashhab_QStatePrep2022, schuld2021qml}. Implementing a \ac{qpie} circuit is equivalent to performing amplitude embedding. The method described by M{\"o}tt{\"o}nen et al.~provides an efficient way to construct such circuits using uniformly controlled rotations~\cite{motteonen2005}.

\section{Classical and Quantum Kernel Methods}
In supervised \ac{ml} problems, kernel methods \cite{koutroumbas2008pattern} are a collection of algorithms commonly used for classification and regression tasks. The core idea of the kernel method is based on the use of feature maps $\phi: X \rightarrow F$ that represent the original data points from the input space $X$ to a higher dimensional feature space $F$, where classification among classes becomes more simple. Kernel methods avoid explicit calculation of the new representation of the point coordinates in the new space $F$ and simply compute the inner products between two data points via a bi-variate function called kernel $K: X \times X \rightarrow \mathbb{R}$ \cite{koutroumbas2008pattern}. Formally, let us consider a classification problem and let $D = \{(\ve{x}_i, y_i)\}_{i=1}^N$ be the dataset, where $\ve{x}_i$ is the data points and $y_i$ their corresponding labels. Then, given a feature map $\phi: X \rightarrow  F$, a kernel function $K$ is written as follows:
\begin{equation}
    K(\ve{x}_{i},\ve{x}_{j}) = \langle \phi(\ve{x}_i), \phi(\ve{x}_j)\rangle,
\end{equation}
where $\langle \cdot, \cdot \rangle$ is the inner product that must satisfy the Mercer condition of positive semi-definiteness \cite{mercer1909xvi,mohri2018foundations} 
\begin{equation}
    \sum_{i=1}^{N} \sum_{j=1}^{N} K(\ve{x}_{i},\ve{x}_{j})c_i c_j \geq 0
\end{equation}
for all choices of real numbers $(c_1,\dots, c_n)$.
This procedure is known as the \textit{kernel trick} and allows us to also work when the data points $\ve{x}$ are not linearly separable. The choice typically depends on the characteristics of the data and the task. 
A straightforward example is the linear kernel, which is given by the formula:
\begin{equation}
K(\ve{x_i}, \ve{x_j}) = \ve{x_i}^T \ve{x_j}.
\end{equation}

\ac{svm} is a well-known kernel-based algorithm \cite{chang2011libsvm, cortes1995support}. The kernel method in \ac{svm} is a powerful technique that enables effective classification by mapping data into a higher-dimensional space, allowing for the identification of an optimal separating hyperplane even in cases where linear separation is infeasible. More precisely, \ac{svm} commonly used in \ac{ml} classification problems with the objective of finding the optimal hyperplane by maximizing the margin between the closest data points of opposing classes. The kernel trick is crucial for constructing the optimal hyperplane by solving the dual optimization problem, which maximizes the function:
\begin{equation}
    L_{D}(\alpha) = \sum_{i=1}^{t} \alpha_{i} - \frac{1}{2} \sum_{i,j=1}^{t} y_{i} y_{j} \alpha_{i} \alpha_{j} K(\ve{x}_{i}, \ve{x}_{j})
    \label{eq:lagrangian}
\end{equation}
with constraints $\sum_{i=1}^{t} \alpha_i y_i = 0$ and $\alpha_i \geq 0$ for all $i$. The solution is given by a nonnegative vector $\boldsymbol{\alpha} = (\alpha_1, \dots, \alpha_t)$. The optimal solution $\boldsymbol{\alpha}^*$ is subsequently employed to construct the classifier: 
\begin{equation}
    m(\ve{s}) = \text{sign}\left(\sum_{i=1}^t y_i\alpha_i^*K(\ve{x}_i,\ve{s})+b\right),
\end{equation}
where $\ve{s}$ is a datum of test set \cite{havlivcek2019supervised} and $b$ is bias. 

In the context of quantum computing, the kernel is constructed using quantum circuits to enhance the computational efficiency of classification using \ac{svm} \cite{havlivcek2019supervised, Khosrojerdi_2025}. Specifically, the quantum kernel estimation involves calculating the overlap between quantum states corresponding to the training data points \cite{schuld2021qml}. To start, we consider zero states $\ket{00\dots0}$ of the $N$-dimensional Hilbert space $\mathcal{H}$, and then we  apply an operator $U_{\phi}(\ve{x})$ over these states and map the classical data point $\ve{x}$ to a vector in the Hilbert space. More precisely, the quantum state that encodes the data point $\ve{x}$ is given by: 
\begin{equation}\label{eq:kernel_pure}
    \ket{\phi(\ve{x})} =U_{\phi}{(\ve{x})}\ket{00\dots0},
\end{equation}
and the quantum kernel between two different pure quantum states is computed by:
\begin{align}
  \nonumber
    K(\ve{x}_i, \ve{x}_j) &= |\bra{0\dots00} U^{\dagger}_{\phi}(\ve{x}_i)U_{\phi}(\ve{x}_j)\ket{00\dots0}|^2 \\
    & =  |\braket{\phi(\ve{x}_{i})}{\phi(\ve{x}_{j})}|^{2},
    \label{eq:kernel}
\end{align}
where $\ve{x}_i$ and $\ve{x}_j$ are two classical data points.

\subsection{Compression-accuracy trade-offs} 

Before presenting numerical results, we first discuss some theoretical arguments that directly 
follow from the information theoretic analysis of Refs.~\cite{banchi2021generalization,banchi2024statistical}.
First of all, from the kernel it is possible to compute the average information content of an ensemble of quantum states.
For instance, let $\{\ket{\phi(\ve{x}_i)}\}_{i=1}^N$ be the ensemble of training states, which can also be 
expressed as a density matrix 
\begin{equation}
  \rho= \frac1N\sum_{i=1}^N \ket{\phi(\ve{x}_i)}\!\!\bra{\phi(\ve{x}_i)}.
  \label{eq:rho}
\end{equation}
The information content of such an ensemble, which is directly linked to its compressibility, can be quantified using either Von Neumann or R\'enyi entropies  $S_\gamma(\rho) = \frac{1}{1-\gamma}\log_2\Tr[\rho^\gamma]$. 
The latter can also be expressed \cite{banchi2021generalization} as $S_\gamma(\rho) = S_\gamma(K)$ where the ``density matrix'' $K$ has elements $K_{ij} = K(\ve{x}_i, \ve{x}_j)/N $ and satisfies the usual properties: $K\geq0$ and $\Tr K=1$. 

Thanks to such result, the entropy of the ensemble \cref{eq:rho} is equal to the entropy of the normalized kernel matrix \cref{eq:kernel}, which does not directly depend on the dimensionality of the quantum Hilbert space, though it is upper bounded by it.
While the number of qubits in the embedding \cref{eq:general_form_qimagestate} quantifies the memory requirement of a single image, the entropy quantifies the information content of an ensemble. 
Large entropy, which results in reduced compression and worse performance \cite{banchi2021generalization},  is possible when the kernel matrix is close to an identity matrix, namely when the off-diagonal elements become small, which happens when the different states $\ket{\phi(\ve{x}_i)}$, $\ket{\phi(\ve{x}_j)}$ become almost orthogonal for $i\neq j$. Therefore, the non-orthogonality of the quantum embedding states, which comes from storing the different pixels via quantum superposition, is the central ingredient behind the enhanced memory storage. 

From these considerations, we can draw a few conclusions, assuming a link between the entropic quantifiers with different $\gamma$. On one hand, smaller entropy guarantees better learning, since the generalization  error is linked to $S_2(\rho)$ \cite{banchi2021generalization}. 
On the other hand, due to Holevo's theorem \cite{holevo2011probabilistic}, $S_1(\rho)$ provides a limit to the amount of information about the embedded images that can be retrieved via quantum measurements, so a smaller entropy means that the training images are more ``hidden''. Nonetheless, this is not necessarily a problem in learning settings, where the task of the algorithm is not to retrieve the image $\ve{x}_i$, but rather its label $y_i$. As long as the quantum state embedding has enough information about the label $y_i$, accurate prediction is still possible. 

\section{Results}
\begin{figure}[t!]
    \centering
    \includegraphics[width=\linewidth]{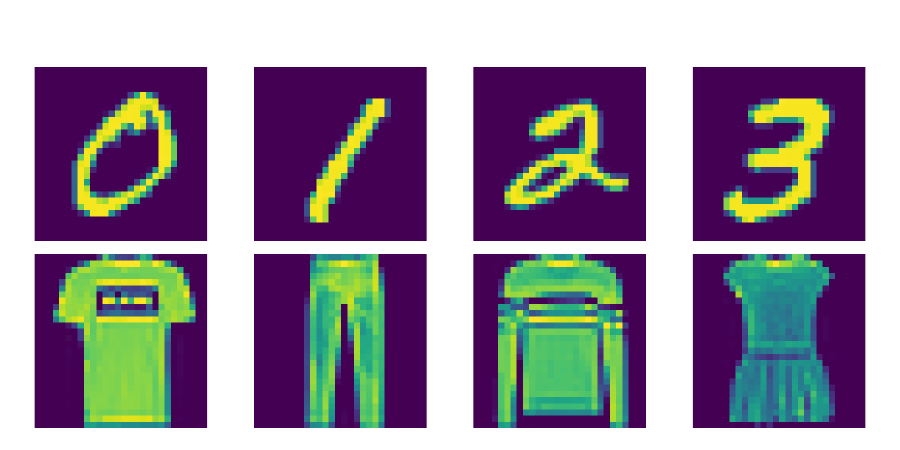}
    \caption{ Image samples of MNIST (top row) and Fashion MNIST (bottom row) datasets.} 
    \label{fig:dataset}
\end{figure}

Simulations are performed using Pennylane~\cite{pennylane} and Scikit-Learn~\cite{scikitlearn}, which are open-source software for quantum and classical machine learning, respectively. 
In particular, Pennylane is used to implement quantum circuits that embed classical images into quantum states, while Scikit-Learn is employed for the SVM-based classification. 
In this study, we adopt MNIST~\cite{lecun2010mnist} and Fashion MNIST~\cite{fmnist2017} datasets, whose samples are illustrated by \cref{fig:dataset}. All images are pre-processed before encoding them into \acp{qimr}. First, the size of each image is reduced to $16 \times 16$ by using the module \emph{cv2.INTER\_AREA} of OpenCV~\cite{opencv}. Secondly, the color of each pixel is resized to $[0, 255]$. 
We perform binary classification using a \ac{svm}  with 
training and testing sets consisting of $10^3$ and $10^2$ images, respectively. 
A summary describing the simulation parameters is reported in \cref{tab:implementation_details}.

\begin{table}[ht]
\centering
\renewcommand{\arraystretch}{1.2}
\begin{tabular*}{0.45\textwidth}{@{\extracolsep{\fill}}p{25mm}p{55mm}}
\toprule
\textbf{Component} & \textbf{Description} \\
\midrule
Datasets: & 
$10^3$ (training) and $10^2$ (test) images from 
MNIST and Fashion-MNIST datasets, grouped as a
binary classification task (class $0$ vs.\ classes $1$–$9$). 
\\
Preprocessing: & Grayscale images resized to $16\times16$ pixels. \\
Model: & Support Vector Machine (SVM) with a precomputed kernel, implemented using the \texttt{scikit-learn} library with default parameters. \\
Validation: & k-fold cross-validation. \\
\bottomrule
\end{tabular*}
\caption{Implementation details for SVM-based binary classification using quantum image representations kernel and classical linear kernel.}
\label{tab:implementation_details}
\end{table}

Given two classical digital images $I$ and $I'$ encoded on the pure quantum states $\ket{I}$ and $\ket{I'}$, respectively, the kernel operator $K$ used in the ideal case is given by:
\begin{equation}\label{eq: quantum_kernel}
    K_{I,I'}  = |\braket{I}{I'}|^{2},
\end{equation}
where $\braket{\cdot}{\cdot}$ denotes the inner product. In the following we calculate \acp{qimr}-based quantum kernels used in the simulations using  the \ac{svm} model. 

The inner product between two MPS states \cref{eq:MPS_realket} reads
\begin{equation}
  \bra{\psi_\Gamma}\psi_{\Gamma'}\rangle = \sum_{ \{i_k\}, \{\lambda_k\}, \{\beta_k\}} \prod_k  \Gamma_{\lambda_k \lambda_{k+1}}^{(k)i_k*} \Gamma'{}_{\beta_k \beta_{k+1}}^{(k)i_k}
\end{equation}
and can be computed efficiently using standard contraction techniques \cite{gray2021hyper}.
Defining:
\begin{equation}
  E_k = \sum_{i_k} \Gamma_{\lambda_k \lambda_{k+1}}^{(k)i_k*} \Gamma'{}_{\beta_k \beta_{k+1}}^{(k)i_k}, \quad k= 1, 2, \dots, n,
\end{equation}
the TNR quantum kernel between two MPS states can be written in compact form as:
\begin{equation}
    \abs{\operatorname{Tr} \left( E_1 E_2 \cdots E_n \right)}^2.
\end{equation}

Given two \ac{frqi} states \cref{eq:frqi_state}, the inner product between them is the following:
\begin{equation*}\label{eq: frqi_inner}
\begin{split}
\braket{I(\theta)}{I(\theta')} & = \frac{1}{2^{2n}} \sum_{i, j=0}^{2^{2n}-1} \braket{c(\theta_i)}{c(\theta_j)} \otimes \braket{i}{j}\\
& = \frac{1}{2^{2n}} \sum_{i, j=0}^{2^{2n}-1} (\cos\theta_{i}\cos\theta_{i}^{'} + \sin\theta_{i}\sin\theta_{i}^{'}) \\
& = \frac{1}{2^{2n}} \sum_{i=0}^{2^{2n}-1} \cos(\theta_{i} - \theta_{i}^{'}), 
\end{split}
\end{equation*}
where we use the orthonormal property of the basis state, i.e.,\ $\braket{i}{j} = \delta_{i, j}$. Thus, the \ac{frqi} quantum kernel is given by:
\begin{equation}
 K_{I,I'}  = \abs{\frac{1}{2^{2n}} \sum_{i=0}^{2^{2n}-1} \cos(\theta_{i} - \theta_{i}^{'})}^2.
\end{equation}
For two \ac{neqr} states \cref{eq:neqr_state} the inner product reads:
\begin{equation*}\label{eq: neqr_inner}
\begin{split}
\braket{I}{I'} & = \frac{1}{2^{2n}} \sum_{y, y', x, x' =0}^{2^{n}-1} \braket{f'(y,x)}{f'(y',x')} \otimes \braket{yx}{y'x'} \\
& = \frac{1}{2^{2n}} \sum_{y=0}^{2^{n}-1}\sum_{x=0}^{2^{n}-1} \delta_{f,f'} \delta_{yx,y'x'},
\end{split}
\end{equation*}
where in the last equality, we use the orthonormal property of the basis states of the same state space. Then, a \ac{neqr} quantum kernel is equal to:
\begin{equation}
 K_{I,I'}  = \abs{\frac{1}{2^{2n}} \sum_{y=0}^{2^{n}-1}\sum_{x=0}^{2^{n}-1} \delta_{f,f'} \delta_{yx,y'x'}}^2.
\end{equation}
Finally, let us consider the inner product between two \ac{qpie} states \cref{eq:qpie_state} is the following:
\begin{equation*}\label{eq: qpie_inner}
    \begin{split}
    \braket{I}{I'} &  = \sum_{i, j =0}^{2^{2n}-1} c_{i} c_{j}^{'} \braket{i}{j} \\
    & = \sum_{i=0}^{2^{2n}-1}  c_{i}c_{i}^{'}.
    \end{split}
\end{equation*}
Accordingly, the \ac{qpie} quantum kernel is given by:
\begin{equation}
    K_{I,I'} =  \abs{\sum_{i=0}^{2^{2n}-1}  c_{i}c_{i}^{'}}^2.
\end{equation}
A summary of \ac{qimr}-based quantum kernels used in the simulations by the \ac{svm} model are reported in \cref{tab:quantum_kernel}.

\begin{table*}[ht]
\centering
\renewcommand{\arraystretch}{1.2}
\begin{tabular*}{\textwidth}{@{\extracolsep{\fill}}p{2.5cm}p{8cm}p{5cm}}
\toprule
\textbf{QImRs} & \textbf{Quantum State} & \textbf{Quantum Kernel} \\
\midrule
\ac{tnr} & $\sum_{i's}\sum_{\lambda's} 
    \Gamma_{\lambda_1\lambda_2}^{(1)i_1} \Gamma_{\lambda_2\lambda_3}^{(2)i_2} \cdots \Gamma_{\lambda_n\lambda_1}^{(n)i_n}
    \ket{i_n \cdots i_2, i_1}$ & $\abs{\operatorname{Tr} \left( E_1 E_2 \cdots E_n \right)}^2$ \\
    FRQI & $ \frac{1}{2^n} \sum_{i=0}^{2^{2n}-1} (\cos\theta_{i} \ket{0} + \sin\theta_{i}\ket{1}) \otimes \ket{i}$ & $ \abs{\frac{1}{2^{2n}} \sum_{i=0}^{2^{2n}-1} \cos(\theta_{i} - \theta_{i}^{'}) }^2$ \\
    NEQR & $\frac{1}{2^n} \sum_{y=0}^{2^{n}-1}\sum_{x=0}^{2^{n}-1} \ket{f(y,x)} \otimes \ket{yx}$ & $\abs{\frac{1}{2^{2n}} \sum_{y=0}^{2^{n}-1}\sum_{x=0}^{2^{n}-1} \delta_{f,f'} \delta_{yx,y'x'}}^2$ \\
    QPIE & $\sum_{i=0}^{2^{2n}-1} c_{i} \ket{i}$ & $\abs{\sum_{i=0}^{2^{2n}-1} c_{i}c_{i}^{'}}^2$ \\
\bottomrule
\end{tabular*}
\caption{We report four different quantum image representations with their quantum states and kernels for $2^n \times 2^n$ grayscale images.}
\label{tab:quantum_kernel}
\end{table*}

\begin{table*}[ht]
\centering
\renewcommand{\arraystretch}{1.}
\begin{tabular*}{\textwidth}{@{\extracolsep{\fill}}p{3cm}p{3cm}p{3.5cm}p{3.5cm}}
\toprule
\textbf{Kernels} & \textbf{Storage} & \textbf{Accuracy} & \textbf{Accuracy} \\
                 & \textbf{Cost}    & \textbf{MNIST} & \textbf{Fashion MNIST} \\
\midrule
 \ac{tnr} & 9 qubits &   0.99 $\pm$ 0.04 & 0.95 $\pm 0.02 $ \\
 \ac{frqi} & 9 qubits &  0.99 $\pm$ 0.00 & 0.95 $\pm 0.02$\\
 \ac{neqr} & 16 qubits & 0.99 $\pm$ 0.00 & 0.94 $\pm 0.02$\\
 \ac{qpie} & 8 qubits &  0.99 $\pm$ 0.00 & 0.96 $\pm 0.02$\\
 Linear & 2\,048 bits & 0.99 $ \pm$ 0.00 & 0.96 $\pm 0.02$\\
\bottomrule
\end{tabular*}
\caption{We report the mean and SEM of the accuracy for different quantum and classical kernels for a binary classification between the pair of classes: $(0, 1), (0, 2), \ldots, (0,9)$ of MNIST and Fashion MNIST datasets. The bond dimension for the \ac{tnr} is equal to 6. Additionally, the storage cost for a $16 \times 16$ grayscale image is reported in terms of qubits (quantum case) and bits (classical case). Quantum kernels on average perform binary classification with the same accuracy as the classical linear kernel, but with exponentially lower storage cost.}
\label{tab: accuracy}
\end{table*}

\begin{figure}
\begin{subfigure}{.475\linewidth}
  \includegraphics[width=\linewidth]{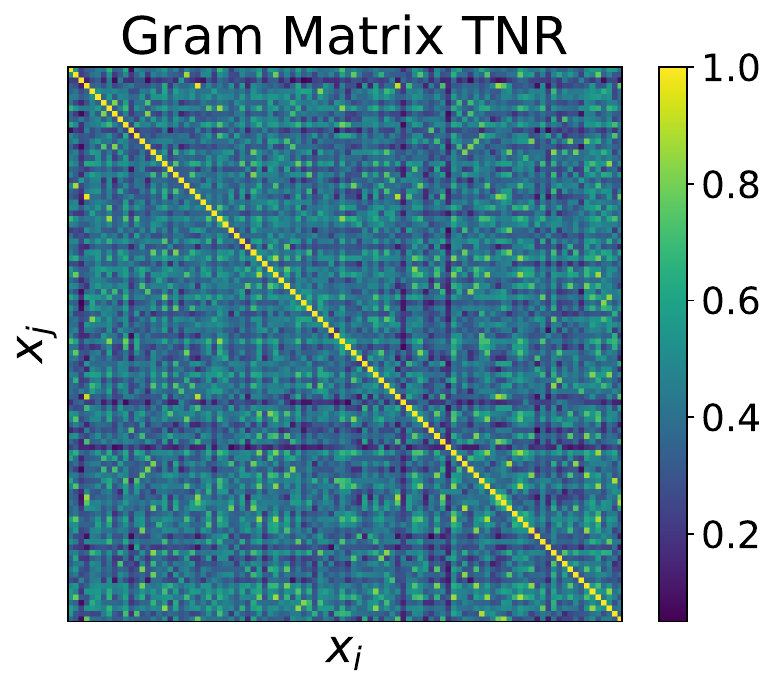}
  \caption{}
  \label{fig:gram_matrix_real_ket}
\end{subfigure}\hfill 
~ 
\begin{subfigure}{.475\linewidth}
  \includegraphics[width=\linewidth]{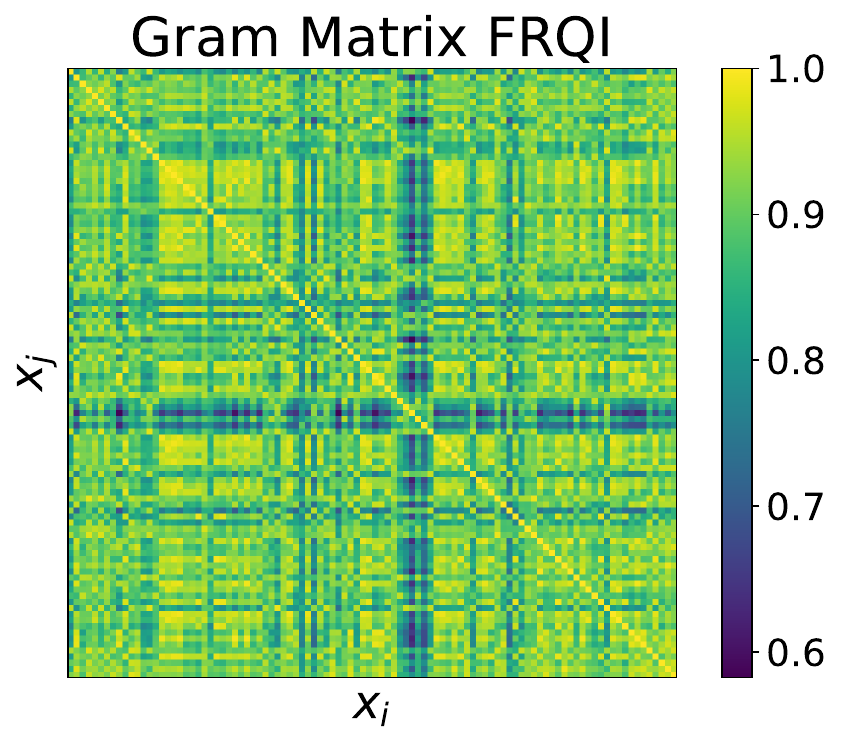}
  \caption{}
  \label{fig:gram_matrix_frqi}
\end{subfigure}

\medskip 
\begin{subfigure}{.475\linewidth}
  \includegraphics[width=\linewidth]{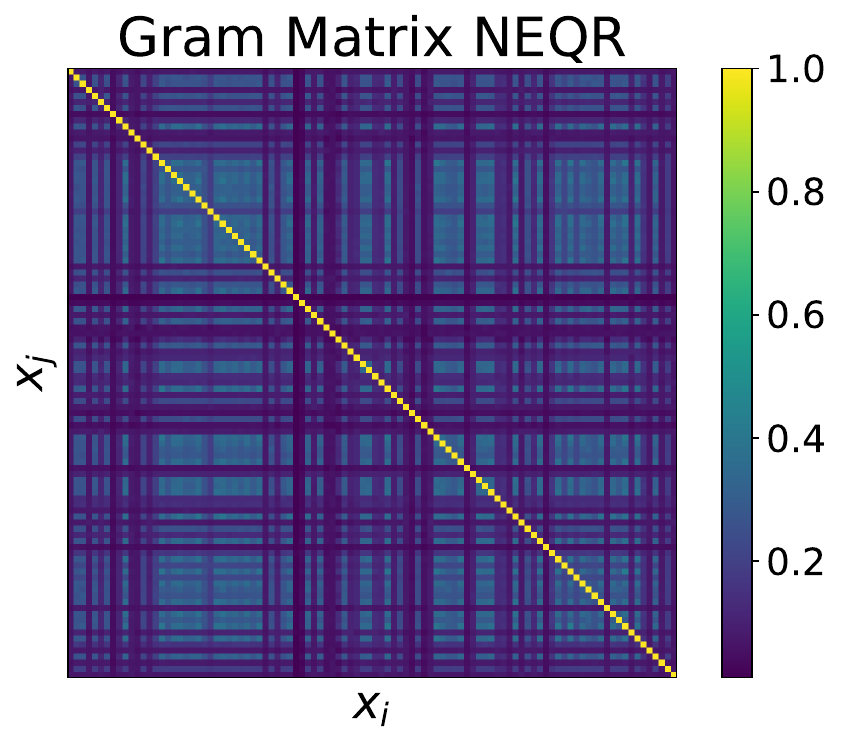}
  \caption{}
  \label{fig:gram_matrix_neqr}
\end{subfigure}\hfill 
\begin{subfigure}{.475\linewidth}
  \includegraphics[width=\linewidth]{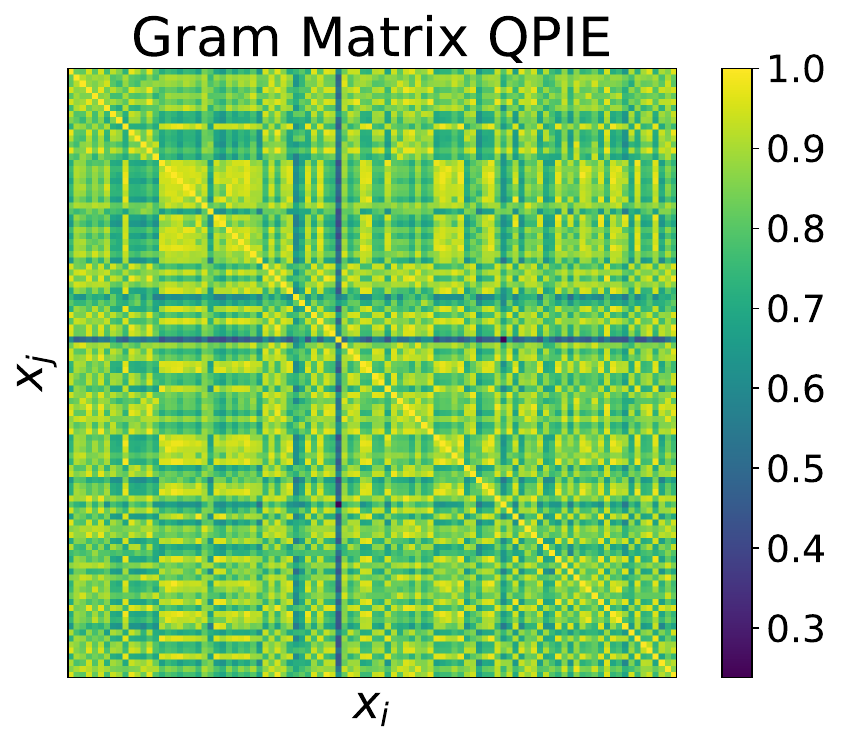}
  \caption{}
  \label{fig:gram_marix_qpie}
\end{subfigure}
\caption{The Gram matrices of 100 training images $x_i$ 
from Fashion MNIST for the quantum kernels: \ac{tnr} (a), \ac{frqi} (b), \ac{neqr} (c) and \ac{qpie} (d). Classes 0 and 1 have been selected. The elements on the diagonal are the kernel of a quantum state $\ket{I}$ with itself, while the elements off the diagonal are the kernel between two different quantum states $\ket{I}$ and $\ket{I'}$. The different off-diagonal coloring of the four Gram matrices, describing the overlap value of two quantum states, gives information on how the different \acp{qimr} compress the classical information into quantum states.}
\label{fig:gram_matrices}
\end{figure}

Successively, to investigate the compression performance of the four \acp{qimr}, we compute the Gram matrices of $100$ training images for two classes $0$ and  $1$, and show them in \cref{fig:gram_matrices}. In fact, the elements of the Gram are the overlap between two quantum states that encode the information of two classical images. In particular, the elements of the Gram matrix take values in the interval $[0, 1]$, where $1$ indicates that the two quantum states completely overlap, while $0$ do not overlap.
The different off-diagonal colorings of the four Gram matrices represent the overlap value between two quantum states encoding different images. From \cref{fig:gram_matrices}, we see that \ac{frqi} and \ac{qpie}  perform a high compression of the classical image information with respect to \ac{neqr} in the encoding procedure. In fact, the most off-diagonal elements of \ac{frqi} and \ac{qpie}  Gram matrices are  in the range $[0.8, 1]$, while for \ac{neqr} case they are in the interval $[0, 0.4]$.  The \ac{tnr} Gram matrix shows an intermediate level of compression with respect to \ac{frqi}, \ac{qpie}, and \ac{neqr}, and its elements are in the range $[0.4, 0.7]$.

We computed the mean and Standard Error on the Mean (SEM) of the accuracy for binary classification between the pair of classes: $(0, 1), (0, 2),\dots, (0,9)$. Accuracy is obtained by the \textit{accuracy\_score} function of the Scikit-Learn~\cite{scikitlearn} library and is computed by the equation:
\begin{equation}
    \text{accuracy} (y, \hat{y}) = \frac{1}{N} \sum_{i=0}^{N-1} 1(\hat{y}_i = y_i),
\end{equation}
where $y_i$ and $\hat{y}_i$ are, respectively, true and the predicted value of the $i$-sample, $N$ is the size of the dataset, and $1(\cdot)$ is the indicator function.
Our findings show that \ac{qimr}-based kernels achieve average accuracy comparable to classical linear kernel on both datasets, while requiring exponentially fewer computational resources to store the image data. The results are reported in \cref{tab: accuracy}.

\begin{figure}[t!]
    \centering
    \includegraphics[width=0.9\linewidth]{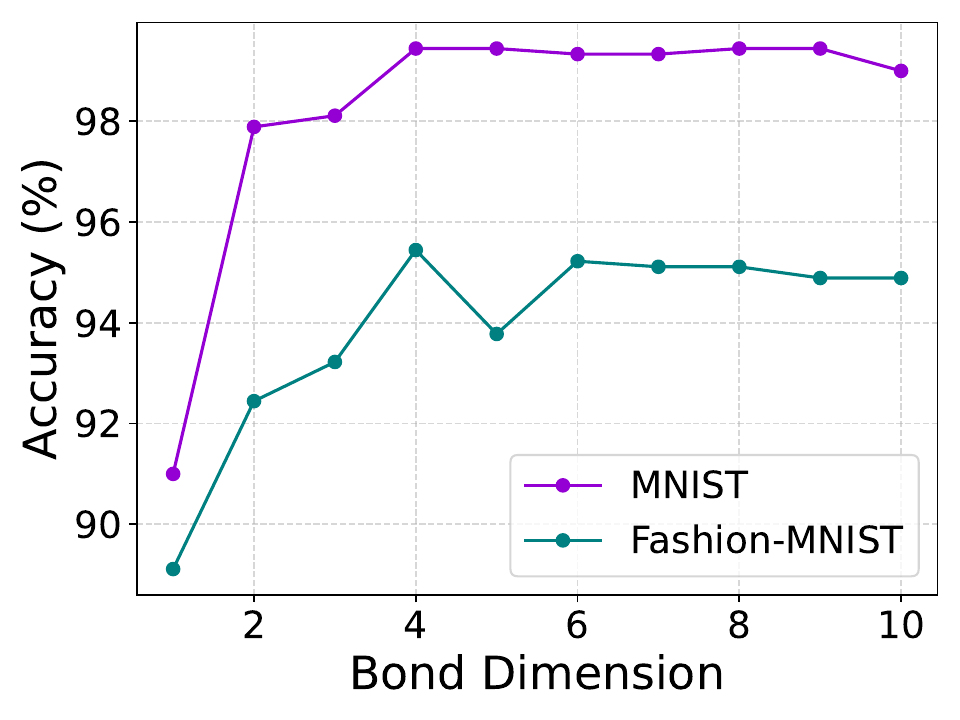}
    \caption{Classification accuracy of the SVM model as a function of bond dimension} 
    \label{fig:acc_vs_bond}
\end{figure}

For the \ac{tnr}, the \ac{svm} model's classification accuracy is evaluated across different bond dimensions for both the MNIST and Fashion-MNIST datasets, as illustrated in  \cref{fig:acc_vs_bond}. The x-axis represents the bond dimension, which governs the level of entanglement captured by the \ac{mps}, while the y-axis shows the classification accuracy (\%) achieved using the \ac{mps}-encoded kernel. The plot suggests that at low bond dimensions, accuracy is limited due to insufficient encoding capacity. However, as the bond dimension increases, the accuracy improves, indicating that the \ac{mps} representation captures richer image features, enhancing classification performance. The model attains its peak accuracy of $[99.4\%]$ on the MNIST dataset and $[95.4\%]$ on the Fashion-MNIST dataset at a bond dimension of four, indicating that moderate bond dimensions are sufficient to capture the essential image correlations. Overall, the results show that the average accuracy for MNIST is consistently higher than for Fashion-MNIST, as expected due to the simpler visual structure of the MNIST digits. Moreover, for MNIST, the accuracy converges once the bond dimension reaches four, whereas for Fashion-MNIST, convergence occurs only after the bond dimension exceeds six.

\section{Discussion}
In this work, we compare different quantum image representation techniques for grayscale digital image to qualitatively analyze the degree of compression when classical image information is encoded in a quantum state and to study the trade-off between accuracy and memory for binary classification problems using kernel methods.

In order to investigate how the four \acp{qimr} compress the classical information of an image, we compute their Gram matrices whose elements are the fidelity (overlap) between two quantum encoding states. We find that with the \ac{frqi} and \ac{qpie} representations, the quantum states are very close to each other after the encoding procedure, while \ac{neqr} shows the opposite scenario. In other words, when the classical information of images is encoded in quantum systems, \ac{frqi} and \ac{qpie} perform higher compression than \ac{neqr}. For  \ac{tnr} encoding, instead we have an intermediate scenario between \ac{frqi} and \ac{neqr}.

Subsequently, we perform image binary classification using quantum kernels based on \ac{qimr} and compare their performance with the classical linear kernel. We find that the mean accuracy of quantum kernels is comparable to the classical one, but quantum kernels require exponentially fewer computational resources to store the image than their classical counterpart. However, we highlight that to date the best known protocols for loading an exact representation of the data into an $n$-qubit state require a number of quantum gates equal to $\mathcal O(2^n)$, which predominates the complexity of quantum algorithms and compromises its potential quantum advantage \cite{aaronson2015read}. An efficient loading of the classical date into a quantum state is still an open question.
A future research direction could be to extend the study to RGB images and compare them with other types of classical kernels. Finally, another interesting perspective could be the possibility of studying the accuracy performance of \ac{qimr} kernels when they are implemented on quantum hardware and noise effects occur.

\section*{Acknowledgments}
M.P., M.K., F.C., and L.B. acknowledge financial support from: PNRR Ministero Università e Ricerca Project No. PE0000023-NQSTI funded by European Union-Next-Generation EU; 
F.C. also acknowledges financial support from the MUR Progetti di Ricerca di Rilevante Interesse Nazionale (PRIN) Bando 2022 - project n.
20227HSE83 – ThAI-MIA funded by the European Union-Next Generation EU;
L.B. also acknowledges Prin 2022 - DD N.~104 del 2/2/2022, entitled ``understanding the LEarning process of QUantum
Neural networks (LeQun)'', proposal code 2022WHZ5XH, CUP B53D23009530006. 

\section*{Data Availability}
The data that support the findings of this study are available from the corresponding author upon reasonable request.

\section*{Competing Interests}
The authors declare no competing interests.

\bibliography{aipsamp}

\end{document}